\documentclass{article}
\usepackage{ijcai23}

\usepackage{times}
\usepackage{soul}
\usepackage{url}
\usepackage[colorlinks, linkcolor=red, anchorcolor=red, citecolor=cyan, breaklinks]{hyperref}
\usepackage[utf8]{inputenc}
\usepackage[small]{caption}
\usepackage{graphicx}
\usepackage{amsmath}
\usepackage{amsthm}
\usepackage{booktabs}
\usepackage{algorithm}
\usepackage{algorithmic}
\usepackage{amssymb}
\usepackage[switch]{lineno}
\usepackage{multirow}
\usepackage{float}

\title{Enhancing Protein Predictive Models via Proteins Data Augmentation: \\ A Benchmark and New Directions}

\author{
Rui Sun$^{1,2,*}$
\and
Lirong Wu$^{1,3,}$\thanks{: Equal Contribution. $\dagger$: Corresponding Author.}\and
Haitao Lin$^1$\and
Yufei Huang$^1$\and
Stan Z. Li$^{1,\dagger}$
\affiliations
$^1$Westlake University,
$^2$Tongji University,
$^3$Zhejiang University
\emails
jui.sun@outlook.com\\
\{
wulirong, huangyufei, linhaitao, stan.zq.li\}@westlake.edu.cn
}
\begin{document}

\maketitle

\begin{abstract}
Augmentation is an effective alternative to utilize the small amount of labeled protein data. However, most of the existing work focuses on designing new architectures or pre-training tasks, and relatively little work has studied data augmentation for proteins. This paper extends data augmentation techniques previously used for images and texts to proteins and then benchmarks these techniques on a variety of protein-related tasks, providing the first comprehensive evaluation of protein augmentation. Furthermore, we propose two novel semantic-level protein augmentation methods, namely \textit{Integrated Gradients Substitution} and \textit{Back Translation Substitution}, which enable protein semantic-aware augmentation through saliency detection and biological knowledge. Finally, we integrate extended and proposed augmentations into an augmentation pool and propose a simple but effective framework, namely \textit{\underline{A}utomated \underline{P}rotein \underline{A}ugmentation} (APA), which can adaptively select the most suitable augmentation combinations for different tasks. Extensive experiments have shown that APA enhances the performance of five protein-related tasks by an average of 10.55\% across three architectures compared to vanilla implementations without augmentation, highlighting its potential to make a great impact on the field.
\end{abstract}

\vspace{-1.2em}
\section{Introduction}
\vspace{-0.2em}
Proteins, composed of one or several chains of amino acids, are fundamental biological entities that play a crucial role in life activities. With the development of Deep Learning (DL) techniques, data-driven protein modeling has made remarkable progress due to its superior performance in modeling complex non-linear relationships \cite{wu2022survey}. A large number of architectures have been adopted to model protein sequences, including Recurrent Neural Networks (RNNs) \cite{armenteros2020language}, Long Short-Term Memory (LSTM) \cite{hochreiter1997long}, ResNet \cite{he2016deep}, and Transformer \cite{vaswani2017attention}. On the other hand, the scarcity of labeled protein data leads to tremendous efforts on protein pre-training \cite{wu2022survey}, whose primary goal is to extract transferable knowledge from massive unlabeled data and then generalize it to various protein-related applications \cite{wu2021self}. Representative tasks used for pre-training protein sequences include Masked Language Modeling (MLM) \cite{rives2021biological,elnaggar2020prottrans,wu2024mapeppi}, Contrastive Predictive Coding (CPC) \cite{lu2020self,wu2024protein}, and Next Amino acid Prediction (NAP) \cite{alley2019unified}, etc. 

High-quality data annotation for the life sciences is especially expensive and time-consuming, leading to the number of labels not satisfying the data-driven methods training. There have been many advances in protein pre-training to utilize \emph{unlabeled data}. Another research interests focus on how to augment those \emph{labeled data} to improve generalization performance. Data augmentation has been widely used and extensively studied for images, texts and graphs \cite{wu2022knowledge,wu2022graphmixup}. For example,   flipping, cropping, rotating, panning, and mixup \cite{shorten2019survey} are representative image augmentation techniques. However, augmentation methods for protein data are still not fully explored. Therefore, in this paper, we study the problem of protein (sequence) augmentation for protein-related applications, which aims to generate more training data by creating plausible variations of existing labeled data without additional ground-truth labels. \textbf{We collect and benchmark existing text and image augmentation techniques, transfer them into protein augmentation methods, and establish a complete pool of protein augmentation that can be categorized into token-level, sequence-level, and semantic-level.}

The benchmark results showed that applying image and text augmentation techniques directly to proteins may be suboptimal. \textit{Firstly}, protein sequences have biological semantic information, which is vital for performing specific functions. \textit{Secondly}, existing text augmentation methods (e.g., synonym substitution) are hard to apply to protein sequences because there is no protein sequence dictionary. To address the problems, \textbf{we propose two semantic-level augmentation methods for proteins: \textit{Integrated Gradients Substitution} and \textit{Back Translation Substitution}}. The former is to discover the saliency regions of the input for augmentation, and the latter enables bio-inspired augmentation through bidirectional translation of protein sequences and mRNA sequences. 

With the extended and proposed augmentation methods, selecting augmentation methods for different datasets, tasks, and architectures becomes an important issue. In practice, a combination of augmentation methods is often more effective than a single augmentation, so we need an appropriate strategy for combining augmentation methods. To tackle this problem, \textbf{we propose a simple but effective framework,  \textit{\underline{A}utomated \underline{P}rotein \underline{A}ugmentation} (APA)}, which adaptively selects the most suitable augmentation combinations for different tasks based on the validation accuracy. Extensive experiments have shown that APA improves five protein-related tasks by an average of 10.55\% across three architectures compared to vanilla implementations without augmentation.

\vspace{0.2em}
Our main contributions can be summarized as follows:
\vspace{-0.2em}
\begin{itemize}
    \item We extend existing augmentation techniques previously used for images and texts to protein data and benchmark them on various protein-related tasks.
    \item We propose two semantic-level protein augmentations: \textit{Integrated Gradients Substitution} and \textit{Back Translation Substitution}, both of which aim to perturb the protein sequences without changing the protein semantics.
    \item We integrate all extended and proposed augmentation methods into a pool of protein augmentation and propose an \textit{\underline{A}utomated \underline{P}rotein \underline{A}ugmentation} (APA) framework that adaptively selects the most suitable augmentation combinations for different architectures and tasks.
\end{itemize}

\vspace{-0.8em}
\section{Related Work}
\textbf{Data Aumentation for Images and Texts.}
Data augmentation has been extensively studied for image and text data. The image augmentation consists of basic geometric augmentation, AutoMix\cite{liu2022automix}, OpenMixup\cite{liu2023harnessing}, random erasing \cite{zhong2020random}, adversarial training \cite{li2018learning}, feature augmentations \cite{devries2017dataset}, etc. In the text domain, representative data augmentation methods include Easy Data Augmentation (EDA) \cite{wei2019eda}, SeqMix \cite{guo2020sequence}, contextual augmentation \cite{kobayashi2018contextual}, BackTranslation \cite{xu2011improving}, etc. Due to space limitations, we refer interested readers to a recent survey \cite{feng2021survey}.
\newline

\vspace{-0.8em}
\noindent\textbf{Protein Pre-training.}
Protein pre-training involves using large-scale unlabeled protein data to pre-train the model and fine-tune it for specific downstream tasks, which can be broadly categorized into unsupervised and semi-supervised methods. The unsupervised pre-training enables the use of large numbers of unlabeled data to obtain protein representations that are more informative for downstream tasks, including ESM \cite{rives2021biological}, GearNet \cite{zhang2022protein}, etc. The semi-supervised pre-training guides the learning of many protein sequences with positional functions through a small number of protein sequences with known functions, such as TAPE \cite{rao2019evaluating}. There is a fundamental difference between protein pre-training and protein augmentation; pre-training aims to learn general knowledge from a large amount of \emph{unlabeled data}, whereas augmentation aims to improve the model generalization by creating variations of existing \emph{labeled data}.
\newline

\vspace{-0.8em}
\noindent\textbf{Automated Augmentation}
mainly aims to automatically apply effective data augmentation strategies to improve model generalization and reduce overfitting. Existing techniques mainly explore the optimal combination of data transformations through Bayesian optimization, reinforcement learning, and other algorithms to increase the diversity of datasets without human intervention. Representative works include AutoAugment \cite{cubuk:autoaugment}, RandAugment \cite{cubuk:randaugment}, Fast AutoAugment \cite{lim2019fast}, etc. However, while automated augmentation for image data has been widely studied, it still remains unexplored for proteins.


\vspace{-0.5em}
\section{Methodology}
This section details how to extend image and text augmentation to proteins. Next, we propose two new semantic-level augmentation transformations for proteins and then integrate the above augmentations into a complete pool of protein augmentation. Finally, we introduce our Automated Protein Augmentation framework, automatically searching for the optimal augmentation combination policy.

\begin{figure*}[htb]
\vspace{-2em}
\begin{minipage}{0.24\linewidth}
    \vspace{3pt}
    \centerline{\includegraphics[width=\textwidth]{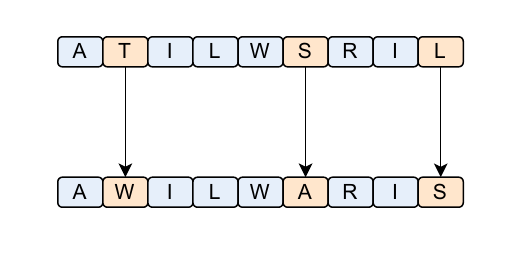}}
    \centerline{(a) Random Substitution}
\end{minipage}
\begin{minipage}{0.24\linewidth}
    \vspace{3pt}
    \centerline{\includegraphics[width=\textwidth]{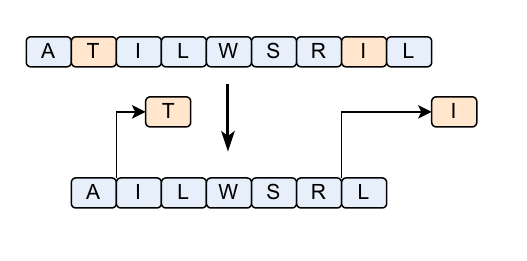}}
    \centerline{(b) Random Deletion}
\end{minipage}
\begin{minipage}{0.24\linewidth}
    \vspace{3pt}
    \centerline{\includegraphics[width=\textwidth]{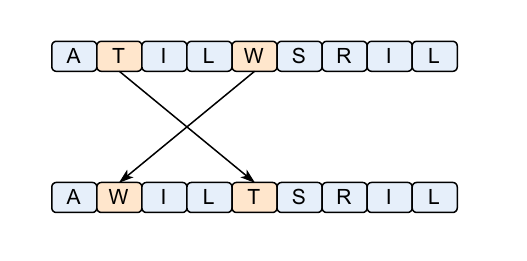}}
    \centerline{(c) Random Swap}
\end{minipage}
\begin{minipage}{0.24\linewidth}
    \vspace{3pt}
    \centerline{\includegraphics[width=\textwidth]{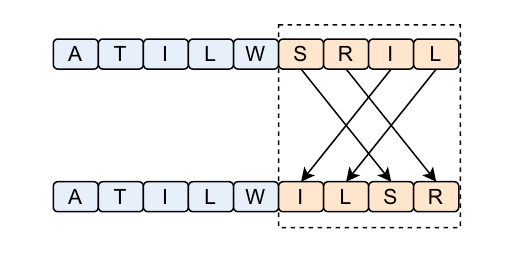}}
    \centerline{(d) Random Shuffle}
\end{minipage}
\begin{minipage}{0.24\linewidth}
    \vspace{3pt}
    \centerline{\includegraphics[width=\textwidth]{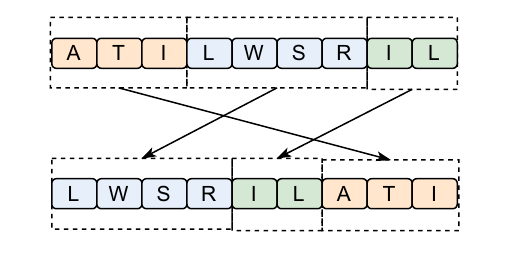}}
    \centerline{(e) Random Cut}
\end{minipage}
\begin{minipage}{0.24\linewidth}
    \vspace{3pt}
    \centerline{\includegraphics[width=\textwidth]{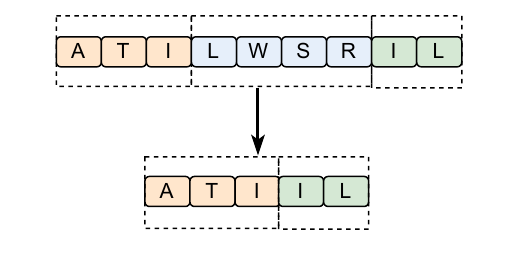}}
    \centerline{(f) Random Subsequence}
\end{minipage}
\begin{minipage}{0.24\linewidth}
    \vspace{3pt}
    \centerline{\includegraphics[width=\textwidth]{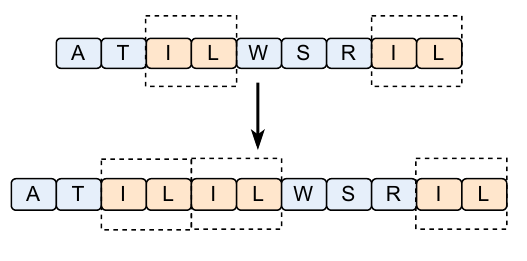}}
    \centerline{(g) Repeat Expansion}
\end{minipage}
\begin{minipage}{0.24\linewidth}
    \vspace{3pt}
    \centerline{\includegraphics[width=\textwidth]{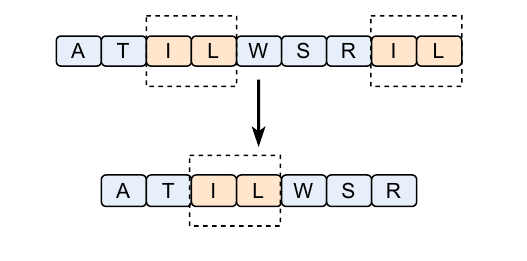}}
    \centerline{(h) Repeat Contraction}
\end{minipage}
\caption{Illustrations of eight protein augmentations, where the unmodified and modified amino acids are marked in blue and orange/green.} \vspace{-0.5em}
\label{figure-baseline}
\end{figure*}

\vspace{-0.5em}
\subsection{Augmentation Extension} \label{sec:3.1}
Our paper mainly focuses on augmentation methods for protein sequence data. We first extend existing image and text augmentations to protein data, which can be divided into token-level and sequence-level categories. A high-level illustration of some augmentations is shown in Figure.~\ref{figure-baseline}; others can be found in \textbf{Appendix A} due to limited limitations.
\subsubsection{Token-Level Augmentation Transformation}
\textit{Random Insertion, Random Substitution, Random Deletion}, and \textit{Random Swap} are four token-level augmentation transformations that are directly extended from EDA \cite{wei2019eda}. Unlike EDA, which uses words as tokens, these methods use individual amino acids as tokens. For \textit{Random Substitution}, we adopt the random replacement strategy, considering that the unique functionality of the 20 amino acids does not allow them to have proximity relationships similar to those of the words in the text.
\subsubsection{Sequence-Level Augmentation Transformation}
For sequence-level augmentation, we augment the sequence or subsequence as a unit, which includes eight methods:
\begin{itemize}
    \item \textit{Random Crop} is similar to \textit{Random Deletion} in token-level, but a subsequence is randomly deleted.
    \item \textit{Global Reverse} is to reverse the entire protein sequence.
    \item \textit{Random Shuffle} is extended from \cite{misra2016shuffle}, which randomly picks a subsequence within a protein and disrupts amino acids within the subsequence.
    \item \textit{Random Cut} and \textit{Random Subsequence} are extended from \cite{takahashi2019data}. \textit{Random Cut} cuts a protein into several sub-sequences and then randomly re-assembles them to form a new protein sequence of the same length. In contrast, \textit{Random Subsequence} only selects some of the sub-sequences and then sequentially assembles them into a new protein sequence.
    

    \item \textit{Repeat Expansion} and \textit{Repeat Contraction} are inspired by AptaTrans \cite{shin2023aptatrans}. In this paper, we use a simple algorithm to find the Frequent Consecutive Subsequence (FCS). \textit{Repeat Expansion} identifies the FCS in the protein sequence and then expands the sequence at a specific location. Conversely, \textit{Repeat Contraction} also locates the FCS but directly deletes some of them.
\end{itemize}
    
We have benchmarked these methods on different protein-related tasks, and a fair comparison is provided in Table.~\ref{table-comparison}.

\begin{figure*}[htb]
\vspace{-1em}
\begin{minipage}[b]{0.3\linewidth}
    \vspace{3pt}
    \centerline{\includegraphics[width=\textwidth]{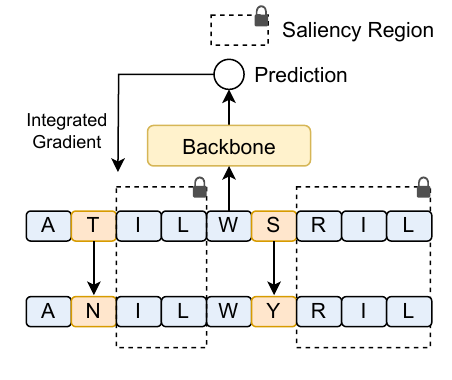}}
    \centerline{(a) Integrated Gradients Substitution}
\end{minipage}
\begin{minipage}[b]{0.7\linewidth}
    \vspace{3pt}
    \centerline{\includegraphics[width=\textwidth]{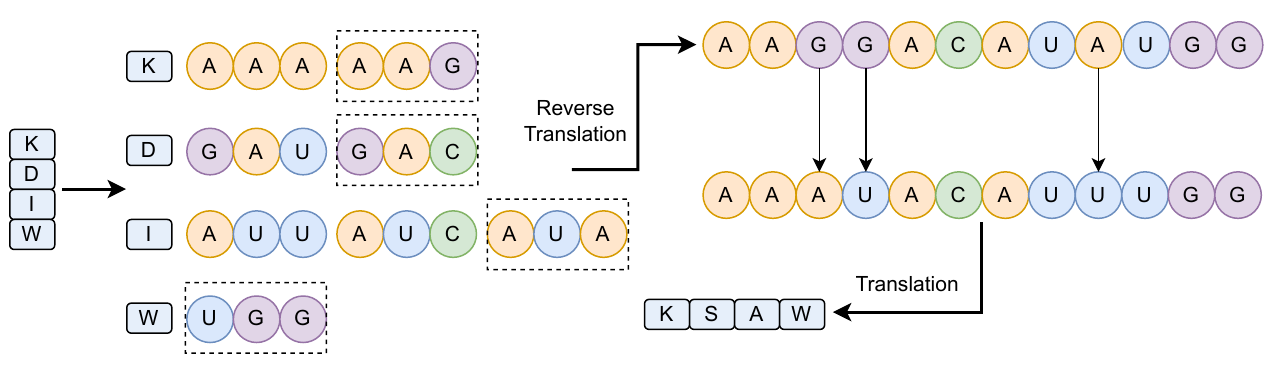}}
    \centerline{(b) Back Translation Substitution}
\end{minipage}
\caption{Illustration of two novel semantic-level augmentation methods. (a) Amino acids within the dashed rectangle represent the saliency regions identified by integrated gradients, with substitutions marked in orange. (b) Rectangles denote amino acids, while circles indicate nucleotides. Given that each amino acid can correspond to multiple codons, it samples the codon for each amino acid, performs reverse translation, introduces random nucleotide substitutions, and then translates the augmented nucleotide sequence back into a protein sequence.} \vspace{-1em}
\label{figure-aug}
\end{figure*}

\subsection{Semantic-Level Protein Transformations} \label{sec:3.2}
We have extended some of the image and text augmentation methods to proteins in Sec.~\ref{sec:3.1}. However, there are still several problems with these methods: (1) Most of the token-level and sequence-level augmentation methods ignore the biological semantics that proteins have. (2) It is hard for one augmentation method to achieve the best results across various protein-related tasks. In other words, these data augmentation methods cannot be adaptively adjusted to different tasks. 

To solve the above two problems, we propose two semantic-level augmentation methods, i.e., \textit{Integrated Gradients Substitution} and \textit{Back Translation Substitution}. A high-level illustration of these two methods is shown in Figure.~\ref{figure-aug}.

\subsubsection{Integrated Gradients (IG) Substitution}
\textit{Integrated Gradients (IG) Substitution} is based on the concept of saliency detection in computer vision \cite{itti1998model}, which is similar to motif discovery in protein modeling. In computer vision, saliency regions of an image contribute significantly to the final perception; similarly, in protein sequences, specific residues or subsequences may greatly impact their function. By identifying these critical regions, our approach not only preserves the biological information of proteins but also can be adapted to different tasks and datasets. We aim to identify these saliency regions within protein sequences by applying the integrated gradients method, as proposed by \cite{sundararajan:axiomatic}. The following formula captures the core of this method:
\begin{align}
\textrm{IG}_i(x)=(x_i-x_i^{\prime})\int_{\alpha=0}^1 \frac{\partial F(x^{\prime}+\alpha(x-x^{\prime}))}{\partial x_i} d\alpha.
\end{align}%

This formula calculates the integrated gradients along the $i^{th}$ dimension of an input $x$ relative to the baseline $x'$. In this context, $F(x)$ denotes the output of the deep learning model for the input $x$, and $\frac{\partial F(x)}{\partial x_i}$ represents the gradient of the model's output w.r.t the $i^{th}$ dimension of input data $x$.
 
This augmentation allows us to pinpoint the residues or subsequences that contribute most significantly to the model's predictions. Once these saliency regions are identified, we preserve them and only modify the other residues during augmentation, ensuring that the sequence's informative segments are maintained, thereby retaining semantic integrity. 

\subsubsection{Back Translation Substitution}
Inspired by the nucleotide sequence augmentation \cite{minot:nucleotide}, we consider that there is also a process of protein translation, from mRNA to protein, in which semantic information is transferred. Thus, we introduce \textit{Back Translation Substitution}, which preserves the original protein semantics by avoiding direct amino acid substitutions and employing a translation and reverse translation process.

We reverse-translate the protein sequence into mRNA, randomly selecting codons due to the codon degeneracy. On this mRNA sequence, we make nucleotide substitutions before translating the sequence back into a protein. The beauty of this approach is its subtlety: it alters the underlying genetic information but does not directly alter the amino acid sequence.This helps preserve the protein sequence's semantic integrity and provides a more subtle augmentation method than direct amino acid substitutions.

\subsection{Automated Protein Augmentation}
\subsubsection{Protein Augmentation Pool Construction}
We detail how to extend existing image and text augmentation methods to proteins and propose two novel semantic-level augmentation methods in Sec.~\ref{sec:3.1} and Sec.~\ref{sec:3.2}, respectively. Next, we construct a pool of protein augmentation that includes all three types of protein augmentation methods: token-level, sequence-level, and semantic-level. The token-level transformations target individual amino acids, treating each amino acid as a discrete token in the protein sequence. Sequence-level transformations involve modifications within specific regions of the sequence, whereas semantic-level transformations are designed to leverage the underlying biological information of the protein sequences.

However, after constructing a complete pool of protein augmentation methods, with both the extended and proposed augmentation methods, how to select the customized augmentations for different datasets, tasks, and architectures becomes an important issue. To address this problem, we propose a simple but effective \textit{\underline{A}utomated \underline{P}rotein \underline{A}ugmentation} (APA) framework that adaptively selects the most suitable augmentation combinations for different tasks.

\subsubsection{Problem Formulation}
We denote the space of the augmentation pool as $\mathbb{O}$ and input data as $\mathcal{X}$. Each augmentation operation $\mathcal{O}\in\mathbb{O}$ is associated with two parameters, i.e., the calling probability $p$ and augmentation magnitude $\lambda$. We denote $S$ be the set of sub-policies where a sub-policy $\tau\in S$ includes $N_\tau$ sequential augmentation operations $\{\bar{\mathcal{O}}_n^{(\tau)}(x;p_n^{(\tau)}, \lambda_n^{(\tau)}):n=1,\ldots,N_{\tau}\}$, where each operations $\bar{\mathcal{O}}$ is defined as follows 
%
\begin{align}
    \bar{\mathcal{O}}(x;p,\lambda):= \left\{
    \begin{aligned}
    &\mathcal{O}(x;\lambda) &&: \textrm{with probability }p\\
    &x &&: \textrm{with probability }1-p \\
    \end{aligned}
    \right.
\end{align}%

\begin{figure*}[htbp]
    \centering
    \includegraphics[width=1\linewidth]{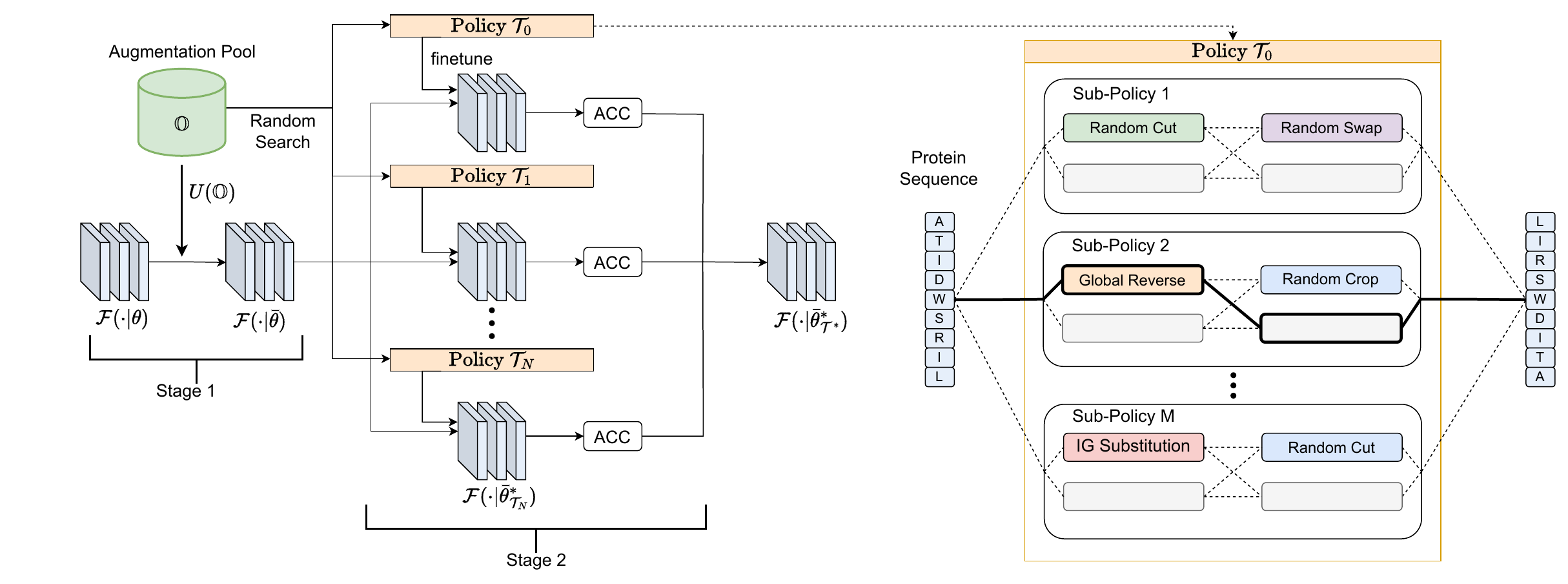}
    \caption{A diagram of Automated Protein Augmentation framework. We first train the initial model with uniform sample policy to obtain the weight-shared model $\mathcal{F}(\cdot|\bar{\theta})$ in stage 1. Then, we fine-tune $\mathcal{F}(\cdot|\bar{\theta})$ on the training set with $N$ different augmentation policies in stage 2, respectively. Finally, we select the best performance policy and fine-tuned model according to the validation accuracy. An illustration of the augmentation policy is shown on the right. Each policy consists of $M$ sub-policies, and each sub-policy has two augmentation transformations sequentially, with two parameters: calling probability $p$ and augmentation magnitude $\lambda$. The right part shows the process of applying a sub-policy to a protein sequence, where the gray vacant rectangle indicates that the transformation is not applied, i.e., $p=0$.}
    \label{fig:framework}
\end{figure*}

A policy $\mathcal{T}$ consists of $M$ sub-policies, as shown in Figure.~\ref{fig:framework}. At each augmentation stage, we randomly apply a sub-policy $\tau$ to training data, instead of the entire policy. For a given model $\mathcal{F}(\cdot|\theta):\mathcal{X}\rightarrow\mathcal{Y}$, parameterized by $\theta$, the model's accuracy and loss on dataset $\mathcal{D}$ are denoted by $\mathcal{R}(\theta|\mathcal{D})$ and $\mathcal{L}(\theta|\mathcal{D})$. The framework performs augmentation on a training dataset  $\mathcal{X}_{\text{train}}$ and evaluates the effectiveness of the augmentation on a validation dataset $\mathcal{X}_{\text{\text{val}}}$. The learning objective is to optimize a policy $\mathcal{T}^{(t)}$ that maximizes the validation accuracy of the model $\mathcal{F}(\cdot|\theta^{t-1})$, which is a typical bi-level optimization problem, defined as follows
\begin{equation}
    \mathcal{T}^{(t)}=\arg \max_{\mathcal{T}} \mathcal{R}\big(\theta^{(t-1)}|\mathcal{T}(\mathcal{X}_{\text{\text{val}}})\big),
\end{equation}
where $\theta^{(t-1)}$ is the parameters of the model $\mathcal{F}(\cdot|\theta)$ trained on the augmented dataset $\mathcal{T}^{(t-1)}(\mathcal{X}_{\text{train}})$, defined as follows
\begin{equation}
    \theta^{(t-1)} = \arg \min_{\mathcal{\theta}} \mathcal{L}\big(\theta|\mathcal{T}^{(t-1)}(\mathcal{X}_{\text{train}})\big).
\end{equation}

\subsubsection{Augmentation Policy Optimization}
The previous subsection showed that the augmentation policy optimization can be formulated as a typical bi-layer optimization problem. In practice, reinforcement learning and Bayesian optimization are commonly used methods for solving it, but they both have some limitations. First, it is worth noting that the search space for augmentation policies (combinations) is huge. In the vast search space, reinforcement learning and Bayesian optimization face tricky efficiency challenges. For example, reinforcement learning requires many environmental interactions for data acquisition and strategy learning, often leading to inefficiency. Besides, Bayesian optimization is based on a Gaussian process, and its computational complexity increases as the number of data points increases, leading to substantial overhead. Furthermore, bi-level optimization requires iterative internal optimization, such as consistent model retraining, which requires extensive computational resources. 

To address the efficiency issue, RandAugment \cite{cubuk:randaugment} found that using a random sampling augmentation strategy can significantly reduce the search space for augmentation while maintaining the same or even better model performance. Besides, AWS \cite{tian:aws} found that data augmentations play a more significant role in the \textbf{later} stages of model training than in the earlier stages. Inspired by these two insights, we propose a simple but effective framework, Automated Protein Augmentation, which divdes the training approach into two stages. In the first stage, we train a weight-shared model $\mathcal{F}(\cdot|\bar{\theta})$ by uniformly sampled augmentations. In the second stage, we fine-tune the weight-shared model with the policy sampled by random search and select the augmentation policy that performs best on the validation set. 

At \textbf{Stage 1}, instead of training from scratch with each augmentation policy, we can focus on the later stages of training. Specifically, in the early stages of training, we only need to train a ``suitable " model $\mathcal{F}(\cdot|\bar{\theta})$ with uniformly sampled augmentations as a basis for fine-tuning the augmentation policy in the second stage, which can significantly reduce the cost of model training. We denote the parameters of the model trained from scratch for the augmentation policy $\mathcal{T}$ as $\theta_{\mathcal{T}}^*$, and the parameters of the model fine-tuned with the weight-shared strategy as $\bar{\theta}_{\mathcal{T}}^*$. According to AWS, they found that the accuracy $R(\bar{\theta}_{\mathcal{T}}^*)$ obtained with uniform sampling is highly correlated with $R(\theta_{\mathcal{T}}^*)$ obtained without the weight-shared strategy. Therefore, uniform sampling is used as the weight-shared strategy. The optimization process of the weight-shared model can be defined as follows:
\begin{equation}
    \bar{\theta}=\arg\min_\theta \frac{1}{|\mathcal{X}_{\text{train}}|}\mathbb{E}_{\mathcal{O}\sim U(\mathbb{O})}\mathcal{L}(\theta|\mathcal{O}(\mathcal{X}_{\text{train}})),
\end{equation}%
where $U(\mathbb{O})$ is the uniform distribution of the protein augmentation methods, and $\mathbb{O}$ is the space of augmentation pool.


At \textbf{Stage 2}, we fine-tune the weight-shared model $\mathcal{F}(\cdot|\bar{\theta})$ with different augmentation policies. The model parameter fine-tuned with augmentation policy $\mathcal{T}_t$ is defined as,
\begin{equation}
    \bar{\theta}^*_{\mathcal{T}_t} = \arg\min_{\theta} \mathcal{L}\big(\bar{\theta} | \mathcal{T}_t (\mathcal{X}_{\text{train}})\big), \forall \  1\leq t \leq N.
\end{equation}%
Next, we select the best performing model on the validation set and its corresponding augmentation policy as follows
\begin{equation}
    \mathcal{T}^* = \arg\max_{\mathcal{T}_t} \mathcal{R}(\bar{\theta}_{\mathcal{T}_t}^* | \mathcal{X}_{\text{val}}).
\end{equation}
Accordingly, we output the model $\mathcal{F}(\cdot|\bar{\theta}^*_{\mathcal{T}^*})$ fine-tuned with the optimal augmentation policy $\mathcal{T}^*$ for prediction.

\begin{table*}
\vspace{-1em}
\centering
\resizebox{0.85\textwidth}{!}{
\begin{tabular}{cc|cccccc}
\toprule
\multicolumn{1}{c}{\textbf{Backbone}} &
  \textbf{Method} &
  \multicolumn{1}{c}{\textbf{EC}} &
  \multicolumn{1}{c}{\textbf{Sub}} &
  \multicolumn{1}{c}{\textbf{Bin}} &
  \multicolumn{1}{c}{\textbf{Fold}} &
  \multicolumn{1}{c}{\textbf{Yst}} &
  \textit{Ave.} \\ \midrule
\multirow{3}{*}{ResNet}    & Vanilla & 0.574$_{\pm0.003}$ & 52.30$_{\pm3.51}$ & 78.99$_{\pm4.41}$ & 8.89$_{\pm1.45}$  & 48.91$_{\pm1.78}$ & - \\
         & w/ APA  & 0.579$_{\pm0.020}$ & 59.09$_{\pm0.65}$ & 86.61$_{\pm0.71}$ & 11.02$_{\pm0.54}$ & 50.59$_{\pm2.37}$ & -       \\
         & $\Delta$ & +0.87\%           & +12.98\%         & +9.65\%          & +23.96\%         & +3.44\%          & +12.51\% \\ \midrule
\multirow{3}{*}{LSTM}      & Vanilla & 0.333$_{\pm0.000}$ & 62.98$_{\pm0.37}$ & 79.74$_{\pm2.06}$ & 8.24$_{\pm1.61}$  & 53.62$_{\pm2.72}$ & - \\
         & w/ APA  & 0.462$_{\pm0.034}$ & 65.95$_{\pm0.30}$ & 88.26$_{\pm0.44}$ & 11.90$_{\pm0.43}$ & 54.63$_{\pm1.95}$ & -       \\
         & $\Delta$ & +38.74\%          & +4.72\%          & +10.68\%         & +44.42\%         & +1.88\%          & +15.43\% \\ \midrule
\multirow{3}{*}{ESM-2-35M} & Vanilla & 0.815$_{\pm0.010}$ & 73.66$_{\pm0.26}$ & 91.13$_{\pm0.35}$ & 28.70$_{\pm0.42}$ & 62.79$_{\pm1.72}$ & - \\
         & w/ APA  & 0.877$_{\pm0.003}$ & 74.88$_{\pm0.50}$ & 92.03$_{\pm0.22}$ & 31.76$_{\pm0.96}$ & 63.72$_{\pm1.56}$ & -       \\
         & $\Delta$ & +7.61\%           & +1.66\%          & +0.99\%          & +10.66\%         & +1.48\%          & +3.70\%  \\ \bottomrule
\end{tabular}}
\caption{\label{citation-guide}
Performance comparison of three backbone models with and without data augmentation using the APA framework on five downstream tasks. The results highlight consistent performance improvements of the APA framework compared to the vanilla implementations.
} \vspace{-0.5em}
\label{table1}
\end{table*}

\vspace{-0.2em}
\section{Experiments}
\subsection{Experimental Setup}
\subsubsection{Protein-related Tasks and Evaluation Metrics}

We evaluate the effectiveness of the APA framework using the PEER benchmark \cite{xu:peer}, which provides a diverse suite of tasks pivotal for understanding protein functions and properties. These tasks include protein function prediction, localization prediction, structure prediction, and protein-protein interaction (PPI) prediction. From these categories, we select five representative protein-related prediction tasks: (1) \textbf{\textit{Enzyme Commission (EC) number prediction}} involves predicting the EC numbers assigned to proteins based on the biochemical reactions they catalyze. (2) \textbf{\textit{Subcellular localization prediction(Subloc)}} is to predict the cellular location of natural proteins, a crucial aspect of understanding their functions and interactions within the cell. (3) \textbf{\textit{Binary Localization Prediction}}, a much simpler version of the Subloc, requires models to categorically classify proteins as either ``membrane-bound" or ``soluble", with binary labels $y\in\{0, 1\}$. (4) \textbf{\textit{Fold Classification}} classifies proteins based on their global structural topology at the fold level, with categorical labels $y \in \{0, 1, \ldots, 1194\}$. The backbone coordinates of each protein structure determine the classification. (5) \textbf{\textit{Yeast PPI prediction}} is to predict whether two yeast proteins interact, represented by a binary label $y\in\{0,1\}$.

For the EC prediction task, we take AUPR as the evaluation metric. For all other four tasks, accuracy is adopted as the evaluation metric. These tasks have been meticulously selected to cover a broad spectrum of protein functions and properties, ensuring a comprehensive evaluation of APA's performance across varied and complex biological scenarios.

\subsubsection{Model Backbones and Prediction Heads}

We follow the guidelines from \cite{xu:peer} to utilize two types of backbone models: protein sequence encoders and pre-trained protein language models. Our protein sequence encoders include two extensively studied models, ResNet and LSTM, as detailed in \cite{rao2019evaluating}. These models offer complementary strengths: ResNet is optimized to capture short-range interactions within protein sequences, focusing on local structural elements. While LSTM is good at capturing long-range interactions, providing deep insights into the overarching sequence architecture. Moreover, complementing these encoders, we also consider a pre-trained protein language model in our framework, specifically ESM-2-35M from \cite{lin2022language}. ESM-2 is a robust model trained on many protein sequences sourced from the UniRef database \cite{suzek2015uniref}. The choice of ESM-2-35M for our evaluation balances training efficiency and performance, making it a suitable candidate for our rigorous protein analysis tasks. This model enriches our framework with advanced capabilities derived from its training on unlabeled protein sequences, offering a deeper understanding of protein functionalities.

Following the backbone, we use different prediction heads that are tailored to the specific requirements of each protein analysis task~\cite{xu:peer}. For tasks related to protein function, localization, and structure prediction, we utilize a 2-layer MLP with ReLU nonlinearity. In the case of Protein-Protein Interaction prediction tasks, we concatenate the embeddings of the two proteins and then apply a 2-layer MLP activated with ReLU to serve as the predictor. Adding a batch normalization layer before the MLP layer is essential for all prediction tasks. This addition standardizes MLP inputs, enhances learning efficiency, and improves model stability.

\subsubsection{Implementation Details}

We employ a cross-entropy loss function for various prediction tasks, such as subcellular localization, binary localization, fold classification, and yeast PPI prediction. Conversely, the enzyme commission task is optimized with a binary cross-entropy loss function. We train each model three times with different random seeds for each task and report the mean and standard deviation. Throughout the training, we conduct ten uniform validation checks, with the best-performing model during these checks being selected for reporting its test performance. Moreover, all experiments are implemented based on the standard implementation using PyTorch 1.8.0 on 4 NVIDIA A100 GPUs. It is worth noting that since Integrated Gradients Substitution is hard to apply to the Yst task, we remove it from the augmentation pool in Table.~\ref{table1}, and do not consider the Yst task in Tables.~\ref{table-comparison}, \ref{table-ablation}. The detailed hyperparameter settings are placed in \textbf{Appendix B}. 


\vspace{-0.2em}
\subsection{Backbone Comparisons within APA}
We report in Table.~\ref{table1} the performance of APA for 5 downstream tasks and 3 architectures and its improvement relative to the vanilla implementation without augmentation, upon analyzing the results in Table.~\ref{table1}, it is evident that the APA framework consistently improves model performance across various downstream tasks. Each backbone model, when paired with APA, outperforms its vanilla counterpart, demonstrating the effectiveness of the augmentation strategies employed by the APA framework. More importantly, the observed improvements are not limited to specific tasks or architectures, and in particular, the improvement of ESM-2-35M suggests that protein augmentation by APA is also helpful for the pre-trained model, implying that protein pre-training and augmentation are complementary.

\begin{table}[!htbp]
\centering
\vspace{0.2em}
\resizebox{0.9\columnwidth}{!}{
\begin{tabular}{l|cccc}
\toprule
\multicolumn{1}{l}{\textbf{Tasks}} & \multicolumn{1}{l}{\textbf{EC}}          & \multicolumn{1}{l}{\textbf{Sub}} & \multicolumn{1}{l}{\textbf{Bin}} & \multicolumn{1}{l}{\textbf{Fold}} \\ \midrule
Vanilla                        & 0.333 & 62.98 & 79.74 & 8.24  \\ \midrule
Random Insertion              & 0.373 & 62.71 & 86.47 & 10.28 \\
Random Substitution           & 0.351 & 64.52 & 82.42 & 10.42 \\
Random Swap                   & 0.280 & 64.99 & 86.42 & 10.69 \\
Random Deletion               & 0.374 & 63.94 & 86.02 & 9.31  \\ \midrule
Random Crop                   & 0.341 & 64.91 & 86.53 & 8.89  \\
Random Shuffle                & 0.346 & 60.88 & 84.70 & 10.14 \\
Global Reverse                & 0.229 & 63.80 & 84.65 & 10.69 \\
Random Subsequence            & 0.313 & 63.94 & 86.59 & 10.00 \\
Random Cut                    & 0.312 & 64.45 & 86.76 & 10.56 \\
Repeat Expansion              & 0.309 & 62.79 & 85.67 & 10.69 \\
Repeat Contraction            & 0.310 & 62.75 & 86.36 & 10.42 \\ \midrule
Back Translation     & 0.360 & 64.73 & 86.82 & 10.97 \\
Integrated Gradients & \underline{0.391} & \underline{65.02} & \underline{87.33} & \underline{11.53} \\ \midrule
APA (ours)            & \textbf{0.462}            & \textbf{65.95}   & \textbf{88.26}     & \textbf{11.90}      \\ \bottomrule
\end{tabular}} 
\caption{\label{citation-guide}
Results of different augmentation operations trained with LSTM for protein prediction on EC, Sub, Bin, and Fold. The \textbf{best} and second scores are marked in \textbf{bold} and \underline{underline}, respectively.} 
\label{table-comparison} \vspace{-1em}
\end{table}

\begin{figure*}[!tbp]
\vspace{-2em}
\centering
\begin{minipage}{0.24\linewidth}
    \centerline{\includegraphics[width=\textwidth]{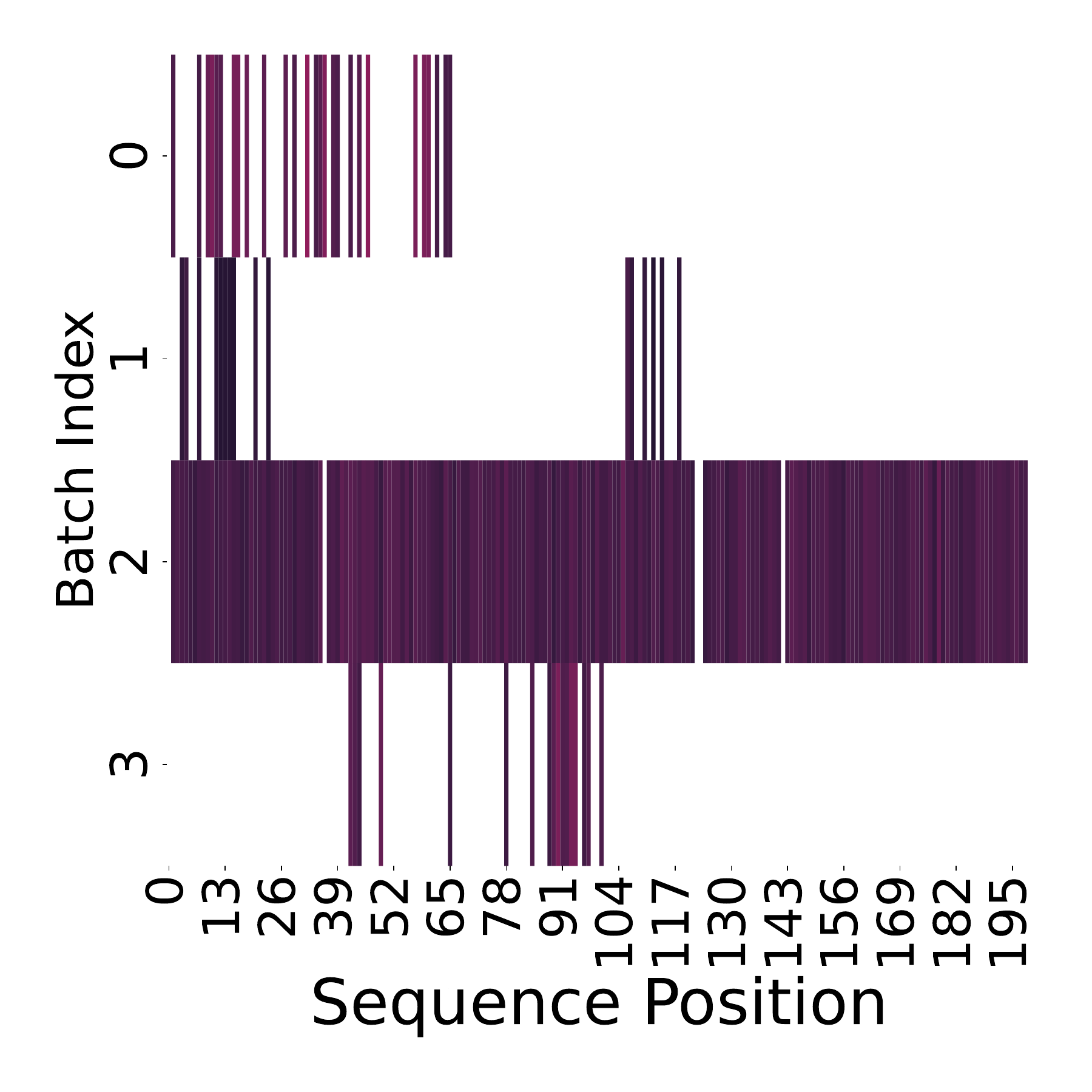}}
\end{minipage}
\begin{minipage}{0.24\linewidth}
    \centerline{\includegraphics[width=\textwidth]{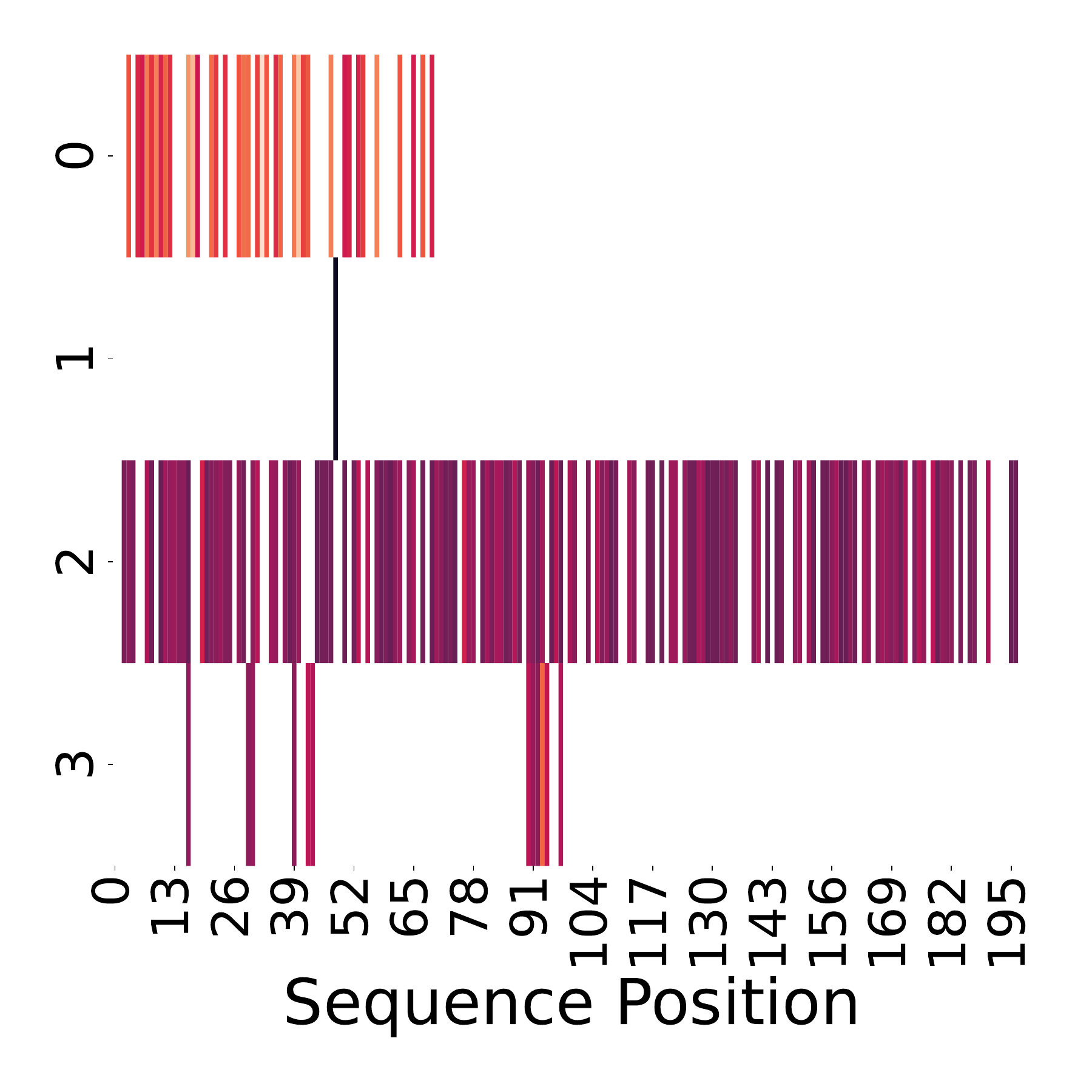}}
\end{minipage}
\begin{minipage}{0.24\linewidth}
    \centerline{\includegraphics[width=\textwidth]{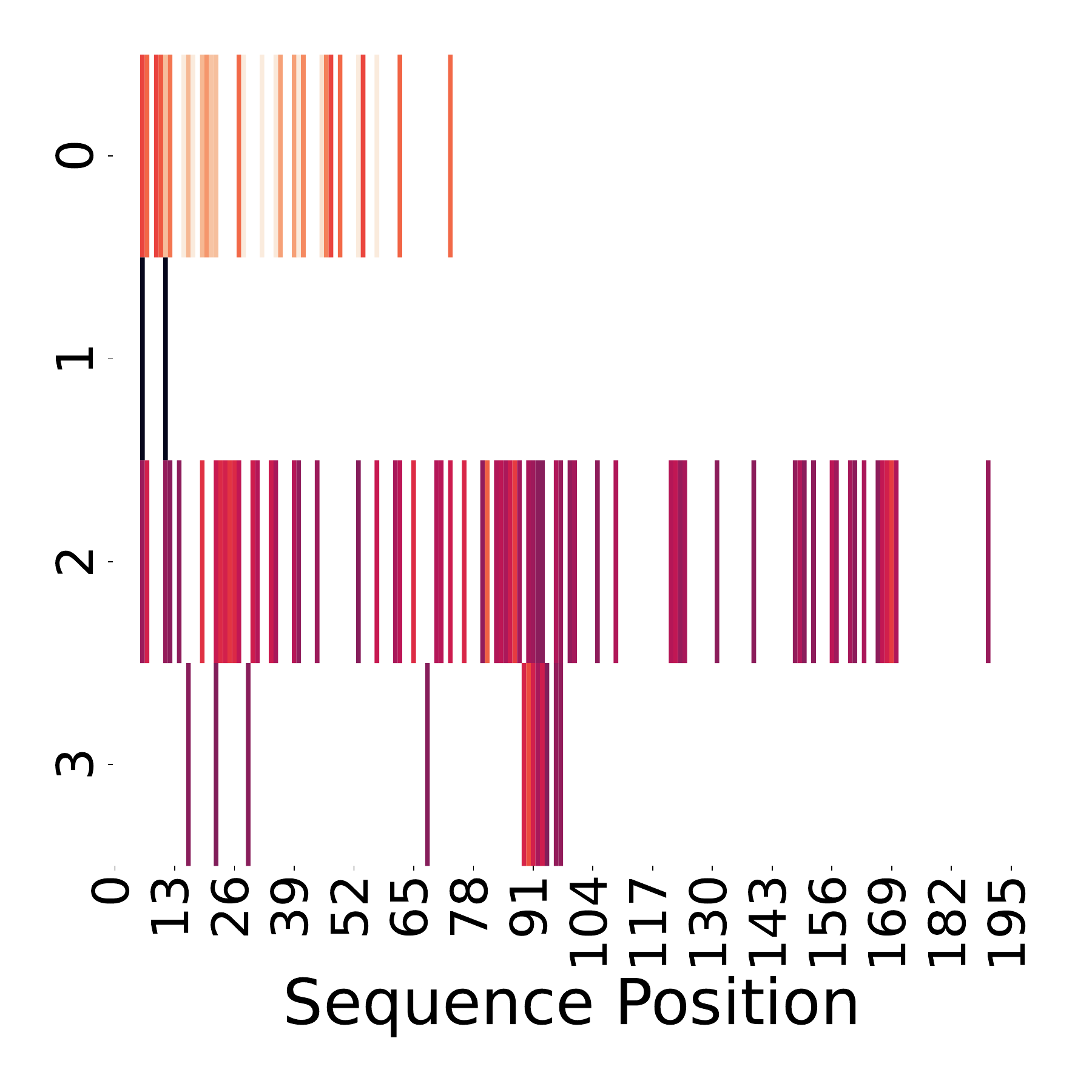}}
\end{minipage}
\begin{minipage}{0.24\linewidth}
    \centerline{\includegraphics[width=\textwidth]{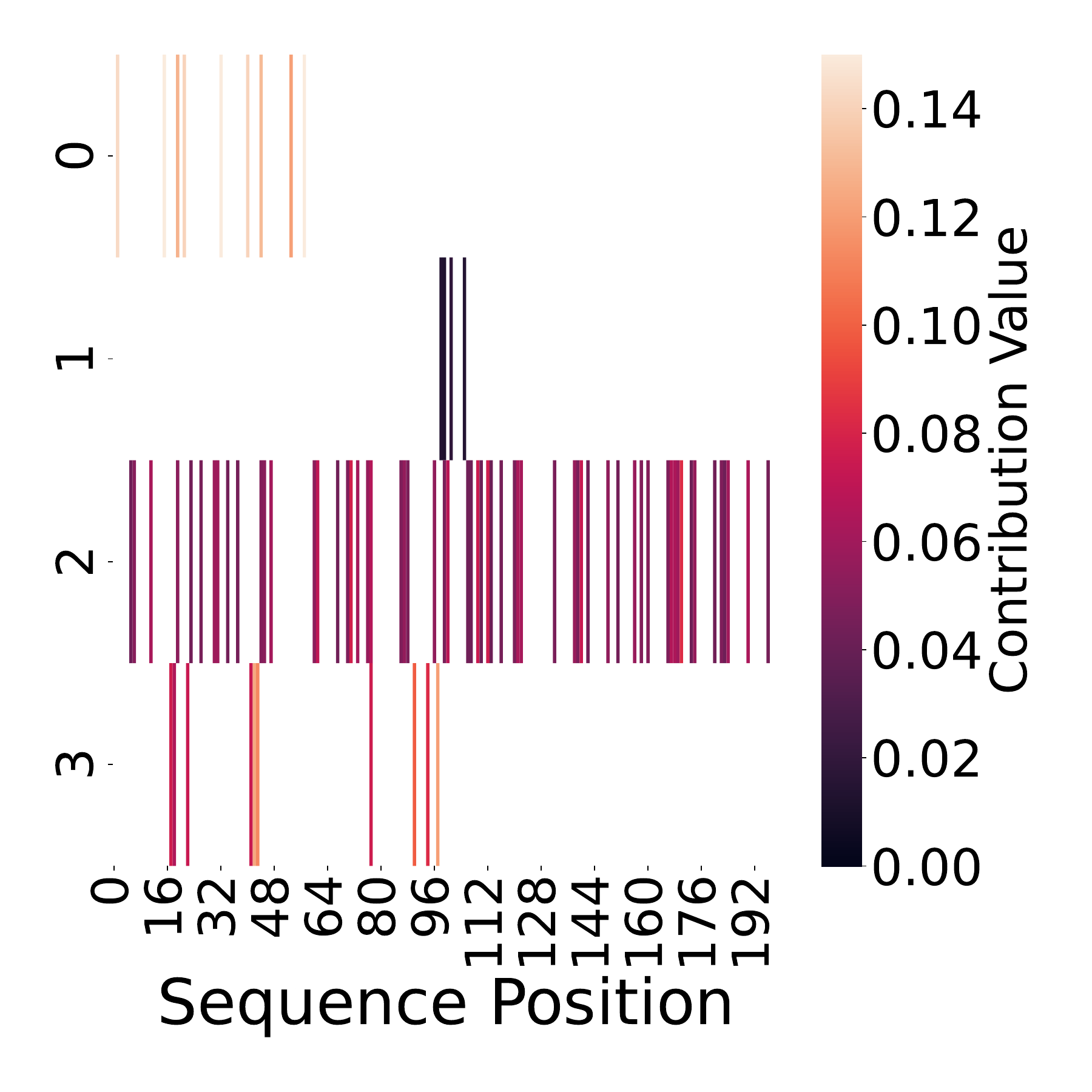}}
\end{minipage}
\caption{Heatmap visualization of Integrated Gradients attributions for 4 proteins in the same batch at epoch 0, 10, 20, and 30 using the ESM-2-35M model on the subcellular localization task. Lighter shades represent regions with higher contributions to the final predictions.} \vspace{-1em}

\label{figure-ig}
\end{figure*}

\subsection{Benchmarks on Protein Augmentation}
We compare the performance of 11 extended augmentations, 2 proposed semantic-level augmentations, and APA on 4 downstream tasks using LSTM as the backbone.
From the results reported in Table.~\ref{table-comparison}, it can be observed that (1) The eleven extended augmentation methods show improved performance over the vanilla implementation in most tasks, which illustrates the importance of protein augmentation. (2) The two semantic-level augmentation operations proposed in this paper, i.e., Back Translation Substitution and Integrated Gradients Substitution, consistently surpass the extended methods across most tasks. These results indicate the effectiveness of these two methods in capturing the complex patterns necessary for accurate protein sequence analysis. (3) Particularly striking is the performance of our APA framework, which outperforms all other baselines. This underscores the strength of our automated augmentation, which combines a variety of data transformations in an adaptive manner, thus improving the model's generalizability.
\vspace{-0.3em}
\subsection{Heatmap Visualization on IG Augmentation}
The visualization presented in Figure.~\ref{figure-ig} provides an interpretive look at the attribution of different regions within protein sequences as understood by the model at different training stages, e.g., epoch 0, 10, 20, and 30. The lighter shades indicate regions with a more significant contribution to the classification outcome. Besides, we showcase only attributions exceeding a 50\% threshold to ensure that the visualization remains clear and emphasizes the most influential factors as perceived by the model.
These heatmaps show that the model's understanding of protein sequences is \emph{dynamic}; the regions deemed prominent in influencing the model's predictions shift as the training progresses. This evolution suggests that it is continually refining its interpretation of the features most essential for accurate classification as it encounters more data. Besides, different protein sequences exhibit different saliency regions, so Integrated Gradients Substitution can achieve adaptive augmentation for various data and tasks.

 

\begin{table}[!tbp]
\centering
\vspace{0.5em}
\resizebox{0.9\columnwidth}{!}{
\begin{tabular}{l|cccc}
\toprule
\textbf{Tasks}           & \textbf{EC}    & \textbf{Sub}   & \textbf{Bin}   & \textbf{Fold}  \\ \midrule
Vanilla         & 0.333 & 62.98 & 79.74 & 8.24  \\ \midrule
Uniform Sample & 0.378 & 63.26 & 84.93 & \underline{11.11} \\
w/o BatchNorm  & 0.179 & 64.73 & \underline{84.99} & 9.17  \\
w/o IG Augmentation         & \underline{0.445} & \underline{65.45} & 83.16 & 10.56 \\ \midrule
APA (full model) & \textbf{0.462} & \textbf{65.95} & \textbf{88.26} & \textbf{11.90} \\ \bottomrule
\end{tabular}}
\caption{\label{citation-guide}
Results of the ablation study show the impact of uniform sampling, the absence of batch normalization, and the exclusion of the Integrated Gradients augmentation on LSTM's performance across four prediction tasks. The decline in performance metrics illustrates the significance of each component in the APA framework.}
\label{table-ablation}
\end{table}
\vspace{-0.5em}
\subsection{Ablation Study}
In the ablation study, the integral components of our APA framework are dissected to understand their individual contributions to model performance. We deconstruct our framework to a simpler uniform sampling process, remove the additional batch normalization layer, and exclude the novel Integrated Gradients (IG) augmentation operation. Table.~\ref{table-ablation} reports a clear degradation in model performance across all tasks upon these modifications, underscoring the importance of each component to the APA framework. A uniform sampling process alone is insufficient to perform, suggesting the need for automated augmentation selection by APA. Besides, removing batch normalization demonstrates its role in stabilizing and improving model training. Furthermore, the absence of the IG operation highlights its value in enhancing the model's predictive capabilities. Together, all these observations validate the well-designed synergies among various components within the proposed APA framework.

\vspace{-0.3em}
\subsection{Training Curve and Early Convergence}
\begin{figure}[!htbp]
    \vspace{-1.2em}
    \centering
    \includegraphics[width=0.9\linewidth]{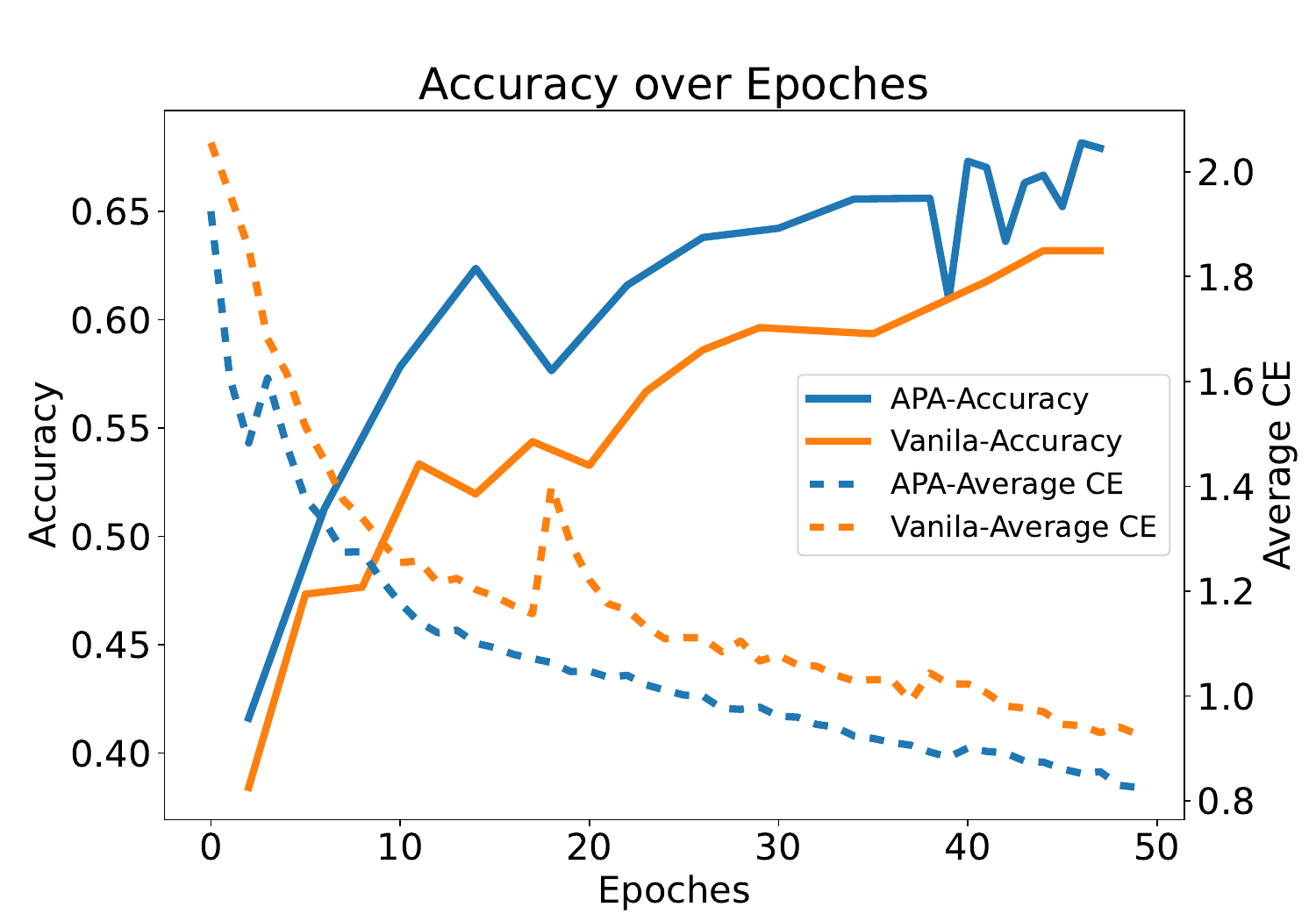}
    \vspace{-0.7em}
    \caption{Comparison of training loss and test accuracy between the vanilla LSTM model and the model with APA over 50 epochs on the Subloc task. The curves demonstrates the advantages of the APA-enhanced model in terms of convergence speed and test accuracy.}
    \label{fig:enter-label}
\end{figure}

We present in Figure.~\ref{fig:enter-label} the training loss (average CE) and test accuracy of the vanilla LSTM model and APA on the subcellular localization task. It can be seen that augmentation by APA can speed up the convergence of the model, thanks to the data diversity captured from the augmented data. More importantly, APA achieves consistent performance gains compared to vanilla implementations across training epochs.

\vspace{-1.5em}
\section{Conclusion}
In this paper, we deeply investigate data augmentation for proteins, a topic that has rarely been explored before. We extend previous augmentation techniques for images and texts to proteins and benchmark them on protein-related tasks. In addition, we propose two semantic-level augmentation methods to leverage protein semantics and biological knowledge. Finally, we construct a pool of protein augmentation to enable architecture- and task-adaptive selection of protein augmentation combinations through an automated protein augmentation framework. Extensive experiments have demonstrated the huge improvement of APA over vanilla implementations for different tasks and architectures. Despite the great progress, limitations still exist; one is the lack of data augmentation for protein structures, and the other is the applicability of APA to pre-trained models with larger amounts of parameters, which are promising directions for future work.

\clearpage




\bibliographystyle{named}
\bibliography{ijcai23}

\begin{thebibliography}{}

\bibitem[\protect\citeauthoryear{Alley \bgroup \em et al.\egroup }{2019}]{alley2019unified}
Ethan~C Alley, Grigory Khimulya, Surojit Biswas, Mohammed AlQuraishi, and George~M Church.
\newblock Unified rational protein engineering with sequence-based deep representation learning.
\newblock {\em Nature methods}, 16(12):1315--1322, 2019.

\bibitem[\protect\citeauthoryear{Armenteros \bgroup \em et al.\egroup }{2020}]{armenteros2020language}
Jose Juan~Almagro Armenteros, Alexander~Rosenberg Johansen, Ole Winther, and Henrik Nielsen.
\newblock Language modelling for biological sequences--curated datasets and baselines.
\newblock {\em BioRxiv}, 2020.

\bibitem[\protect\citeauthoryear{Cubuk \bgroup \em et al.\egroup }{2018}]{cubuk:autoaugment}
Ekin~D Cubuk, Barret Zoph, Dandelion Mane, Vijay Vasudevan, and Quoc~V Le.
\newblock Autoaugment: Learning augmentation policies from data.
\newblock {\em arXiv preprint arXiv:1805.09501}, 2018.

\bibitem[\protect\citeauthoryear{Cubuk \bgroup \em et al.\egroup }{2020}]{cubuk:randaugment}
Ekin~D Cubuk, Barret Zoph, Jonathon Shlens, and Quoc~V Le.
\newblock Randaugment: Practical automated data augmentation with a reduced search space.
\newblock In {\em Proceedings of the IEEE/CVF conference on computer vision and pattern recognition workshops}, pages 702--703, 2020.

\bibitem[\protect\citeauthoryear{DeVries and Taylor}{2017}]{devries2017dataset}
Terrance DeVries and Graham~W Taylor.
\newblock Dataset augmentation in feature space.
\newblock {\em arXiv preprint arXiv:1702.05538}, 2017.

\bibitem[\protect\citeauthoryear{Elnaggar \bgroup \em et al.\egroup }{2020}]{elnaggar2020prottrans}
Ahmed Elnaggar, Michael Heinzinger, Christian Dallago, Ghalia Rihawi, Yu~Wang, Llion Jones, Tom Gibbs, Tamas Feher, Christoph Angerer, Martin Steinegger, et~al.
\newblock Prottrans: towards cracking the language of life's code through self-supervised deep learning and high performance computing.
\newblock {\em arXiv preprint arXiv:2007.06225}, 2020.

\bibitem[\protect\citeauthoryear{Feng \bgroup \em et al.\egroup }{2021}]{feng2021survey}
Steven~Y Feng, Varun Gangal, Jason Wei, Sarath Chandar, Soroush Vosoughi, Teruko Mitamura, and Eduard Hovy.
\newblock A survey of data augmentation approaches for nlp.
\newblock {\em arXiv preprint arXiv:2105.03075}, 2021.

\bibitem[\protect\citeauthoryear{Guo \bgroup \em et al.\egroup }{2020}]{guo2020sequence}
Demi Guo, Yoon Kim, and Alexander~M Rush.
\newblock Sequence-level mixed sample data augmentation.
\newblock {\em arXiv preprint arXiv:2011.09039}, 2020.

\bibitem[\protect\citeauthoryear{He \bgroup \em et al.\egroup }{2016}]{he2016deep}
Kaiming He, Xiangyu Zhang, Shaoqing Ren, and Jian Sun.
\newblock Deep residual learning for image recognition.
\newblock In {\em Proceedings of the IEEE conference on computer vision and pattern recognition}, pages 770--778, 2016.

\bibitem[\protect\citeauthoryear{Hochreiter and Schmidhuber}{1997}]{hochreiter1997long}
Sepp Hochreiter and J{\"u}rgen Schmidhuber.
\newblock Long short-term memory.
\newblock {\em Neural computation}, 9(8):1735--1780, 1997.

\bibitem[\protect\citeauthoryear{Itti \bgroup \em et al.\egroup }{1998}]{itti1998model}
Laurent Itti, Christof Koch, and Ernst Niebur.
\newblock A model of saliency-based visual attention for rapid scene analysis.
\newblock {\em IEEE Transactions on pattern analysis and machine intelligence}, 20(11):1254--1259, 1998.

\bibitem[\protect\citeauthoryear{Keyu \bgroup \em et al.\egroup }{2020}]{tian:aws}
Tian Keyu, Lin Chen, Sun Ming, Zhou Luping, Yan Junjie, and Ouyang Wanli.
\newblock Improving auto-augment via augmentation-wise weight sharing.
\newblock {\em Advances in Neural Information Processing Systems}, 33:19088--19098, June 2020.

\bibitem[\protect\citeauthoryear{Kobayashi}{2018}]{kobayashi2018contextual}
Sosuke Kobayashi.
\newblock Contextual augmentation: Data augmentation by words with paradigmatic relations.
\newblock {\em arXiv preprint arXiv:1805.06201}, 2018.

\bibitem[\protect\citeauthoryear{Li \bgroup \em et al.\egroup }{2018}]{li2018learning}
Shuangtao Li, Yuanke Chen, Yanlin Peng, and Lin Bai.
\newblock Learning more robust features with adversarial training.
\newblock {\em arXiv preprint arXiv:1804.07757}, 2018.

\bibitem[\protect\citeauthoryear{Lim \bgroup \em et al.\egroup }{2019}]{lim2019fast}
Sungbin Lim, Ildoo Kim, Taesup Kim, Chiheon Kim, and Sungwoong Kim.
\newblock Fast autoaugment.
\newblock {\em Advances in Neural Information Processing Systems}, 32, 2019.

\bibitem[\protect\citeauthoryear{Lin \bgroup \em et al.\egroup }{2022}]{lin2022language}
Zeming Lin, Halil Akin, Roshan Rao, Brian Hie, Zhongkai Zhu, Wenting Lu, Nikita Smetanin, Allan dos Santos~Costa, Maryam Fazel-Zarandi, Tom Sercu, Sal Candido, et~al.
\newblock Language models of protein sequences at the scale of evolution enable accurate structure prediction.
\newblock {\em bioRxiv}, 2022.

\bibitem[\protect\citeauthoryear{Liu \bgroup \em et al.\egroup }{2022}]{liu2022automix}
Zicheng Liu, Siyuan Li, Di~Wu, Zihan Liu, Zhiyuan Chen, Lirong Wu, and Stan~Z. Li.
\newblock Automix: Unveiling the power of mixup for stronger classifiers, 2022.

\bibitem[\protect\citeauthoryear{Liu \bgroup \em et al.\egroup }{2023}]{liu2023harnessing}
Zicheng Liu, Siyuan Li, Ge~Wang, Cheng Tan, Lirong Wu, and Stan~Z. Li.
\newblock Harnessing hard mixed samples with decoupled regularizer, 2023.

\bibitem[\protect\citeauthoryear{Lu \bgroup \em et al.\egroup }{2020}]{lu2020self}
Amy~X Lu, Haoran Zhang, Marzyeh Ghassemi, and Alan Moses.
\newblock Self-supervised contrastive learning of protein representations by mutual information maximization.
\newblock {\em BioRxiv}, 2020.

\bibitem[\protect\citeauthoryear{Minot and Reddy}{2023}]{minot:nucleotide}
Mason Minot and Sai~T Reddy.
\newblock Nucleotide augmentation for machine learning-guided protein engineering.
\newblock {\em Bioinformatics Advances}, 3(1):vbac094, 2023.

\bibitem[\protect\citeauthoryear{Misra \bgroup \em et al.\egroup }{2016}]{misra2016shuffle}
Ishan Misra, C~Lawrence Zitnick, and Martial Hebert.
\newblock Shuffle and learn: unsupervised learning using temporal order verification.
\newblock In {\em Computer Vision--ECCV 2016: 14th European Conference, Amsterdam, The Netherlands, October 11--14, 2016, Proceedings, Part I 14}, pages 527--544. Springer, 2016.

\bibitem[\protect\citeauthoryear{Rao \bgroup \em et al.\egroup }{2019}]{rao2019evaluating}
Roshan Rao, Nicholas Bhattacharya, Neil Thomas, Yan Duan, Peter Chen, John Canny, Pieter Abbeel, and Yun Song.
\newblock Evaluating protein transfer learning with tape.
\newblock {\em Advances in neural information processing systems}, 32, 2019.

\bibitem[\protect\citeauthoryear{Rives \bgroup \em et al.\egroup }{2021}]{rives2021biological}
Alexander Rives, Joshua Meier, Tom Sercu, Siddharth Goyal, Zeming Lin, Jason Liu, Demi Guo, Myle Ott, C~Lawrence Zitnick, Jerry Ma, et~al.
\newblock Biological structure and function emerge from scaling unsupervised learning to 250 million protein sequences.
\newblock {\em Proceedings of the National Academy of Sciences}, 118(15):e2016239118, 2021.

\bibitem[\protect\citeauthoryear{Shin \bgroup \em et al.\egroup }{2023}]{shin2023aptatrans}
Incheol Shin, Keumseok Kang, Juseong Kim, Sanghun Sel, Jeonghoon Choi, Jae-Wook Lee, Ho~Young Kang, and Giltae Song.
\newblock Aptatrans: a deep neural network for predicting aptamer-protein interaction using pretrained encoders.
\newblock {\em BMC bioinformatics}, 24(1):447, 2023.

\bibitem[\protect\citeauthoryear{Shorten and Khoshgoftaar}{2019}]{shorten2019survey}
Connor Shorten and Taghi~M Khoshgoftaar.
\newblock A survey on image data augmentation for deep learning.
\newblock {\em Journal of big data}, 6(1):1--48, 2019.

\bibitem[\protect\citeauthoryear{Sundararajan \bgroup \em et al.\egroup }{2017}]{sundararajan:axiomatic}
Mukund Sundararajan, Ankur Taly, and Qiqi Yan.
\newblock Axiomatic attribution for deep networks.
\newblock In {\em International conference on machine learning}, pages 3319--3328. PMLR, 2017.

\bibitem[\protect\citeauthoryear{Suzek \bgroup \em et al.\egroup }{2015}]{suzek2015uniref}
Baris~E Suzek, Yuqi Wang, Hongzhan Huang, Peter~B McGarvey, Cathy~H Wu, and UniProt Consortium.
\newblock Uniref clusters: a comprehensive and scalable alternative for improving sequence similarity searches.
\newblock {\em Bioinformatics}, 31(6):926--932, 2015.

\bibitem[\protect\citeauthoryear{Takahashi \bgroup \em et al.\egroup }{2019}]{takahashi2019data}
Ryo Takahashi, Takashi Matsubara, and Kuniaki Uehara.
\newblock Data augmentation using random image cropping and patching for deep cnns.
\newblock {\em IEEE Transactions on Circuits and Systems for Video Technology}, 30(9):2917--2931, 2019.

\bibitem[\protect\citeauthoryear{Vaswani \bgroup \em et al.\egroup }{2017}]{vaswani2017attention}
Ashish Vaswani, Noam Shazeer, Niki Parmar, Jakob Uszkoreit, Llion Jones, Aidan~N Gomez, {\L}ukasz Kaiser, and Illia Polosukhin.
\newblock Attention is all you need.
\newblock {\em Advances in neural information processing systems}, 30, 2017.

\bibitem[\protect\citeauthoryear{Wei and Zou}{2019}]{wei2019eda}
Jason Wei and Kai Zou.
\newblock Eda: Easy data augmentation techniques for boosting performance on text classification tasks.
\newblock {\em arXiv preprint arXiv:1901.11196}, 2019.

\bibitem[\protect\citeauthoryear{Wu \bgroup \em et al.\egroup }{2021}]{wu2021self}
Lirong Wu, Haitao Lin, Cheng Tan, Zhangyang Gao, and Stan~Z Li.
\newblock Self-supervised learning on graphs: Contrastive, generative, or predictive.
\newblock {\em IEEE Transactions on Knowledge and Data Engineering}, 2021.

\bibitem[\protect\citeauthoryear{Wu \bgroup \em et al.\egroup }{2022a}]{wu2022survey}
Lirong Wu, Yufei Huang, Haitao Lin, and Stan~Z Li.
\newblock A survey on protein representation learning: Retrospect and prospect.
\newblock {\em arXiv preprint arXiv:2301.00813}, 2022.

\bibitem[\protect\citeauthoryear{Wu \bgroup \em et al.\egroup }{2022b}]{wu2022knowledge}
Lirong Wu, Haitao Lin, Yufei Huang, and Stan~Z Li.
\newblock Knowledge distillation improves graph structure augmentation for graph neural networks.
\newblock {\em Advances in Neural Information Processing Systems}, 35:11815--11827, 2022.

\bibitem[\protect\citeauthoryear{Wu \bgroup \em et al.\egroup }{2022c}]{wu2022graphmixup}
Lirong Wu, Jun Xia, Zhangyang Gao, Haitao Lin, Cheng Tan, and Stan~Z Li.
\newblock Graphmixup: Improving class-imbalanced node classification by reinforcement mixup and self-supervised context prediction.
\newblock In {\em Joint European Conference on Machine Learning and Knowledge Discovery in Databases}, pages 519--535. Springer, 2022.

\bibitem[\protect\citeauthoryear{Wu \bgroup \em et al.\egroup }{2024a}]{wu2024protein}
Lirong Wu, Yufei Huang, Cheng Tan, Zhangyang Gao, Haitao Lin, Bozhen Hu, Zicheng Liu, and Stan~Z Li.
\newblock Psc-cpi: Multi-scale protein sequence-structure contrasting for efficient and generalizable compound-protein interaction prediction.
\newblock In {\em Proceedings of the AAAI Conference on Artificial Intelligence}, 2024.

\bibitem[\protect\citeauthoryear{Wu \bgroup \em et al.\egroup }{2024b}]{wu2024mapeppi}
Lirong Wu, Yijun Tian, Yufei Huang, Siyuan Li, Haitao Lin, Nitesh~V Chawla, and Stan Li.
\newblock {MAPE}-{PPI}: Towards effective and efficient protein-protein interaction prediction via microenvironment-aware protein embedding.
\newblock In {\em The Twelfth International Conference on Learning Representations}, 2024.

\bibitem[\protect\citeauthoryear{Xu and Zhang}{2011}]{xu2011improving}
Dong Xu and Yang Zhang.
\newblock Improving the physical realism and structural accuracy of protein models by a two-step atomic-level energy minimization.
\newblock {\em Biophysical journal}, 101(10):2525--2534, 2011.

\bibitem[\protect\citeauthoryear{Xu \bgroup \em et al.\egroup }{2022}]{xu:peer}
Minghao Xu, Zuobai Zhang, Jiarui Lu, Zhaocheng Zhu, Yangtian Zhang, Ma~Chang, Runcheng Liu, and Jian Tang.
\newblock Peer: a comprehensive and multi-task benchmark for protein sequence understanding.
\newblock {\em Advances in Neural Information Processing Systems}, 35:35156--35173, 2022.

\bibitem[\protect\citeauthoryear{Zhang \bgroup \em et al.\egroup }{2022}]{zhang2022protein}
Zuobai Zhang, Minghao Xu, Arian Jamasb, Vijil Chenthamarakshan, Aurelie Lozano, Payel Das, and Jian Tang.
\newblock Protein representation learning by geometric structure pretraining.
\newblock {\em arXiv preprint arXiv:2203.06125}, 2022.

\bibitem[\protect\citeauthoryear{Zhong \bgroup \em et al.\egroup }{2020}]{zhong2020random}
Zhun Zhong, Liang Zheng, Guoliang Kang, Shaozi Li, and Yi~Yang.
\newblock Random erasing data augmentation.
\newblock In {\em Proceedings of the AAAI conference on artificial intelligence}, volume~34, pages 13001--13008, 2020.

\end{thebibliography}

\renewcommand\thefigure{A\arabic{figure}}
\renewcommand\thetable{A\arabic{table}}
\setcounter{table}{0}
\setcounter{figure}{0}

\clearpage
\appendix
\section{Augmentation Extension Illustration}
Figure.~\ref{figure-appendix-extended} shows augmentation methods that were not shown in Section~\ref{sec:3.1}. They, along with the eight augmentation methods mentioned above, constitute the token-level and sequence-level categorizations in the augmentation pool.

\begin{minipage}{0.45\linewidth}
    \vspace{3pt}
    \centerline{\includegraphics[width=\textwidth]{figures/Random-Substitution.drawio.pdf}}
    \centerline{(a) Random Insertion}
\end{minipage}
\begin{minipage}{0.45\linewidth}
    \vspace{3pt}
    \centerline{\includegraphics[width=\textwidth]{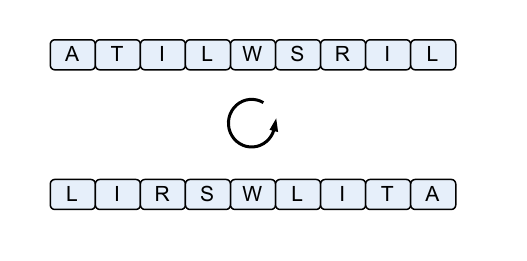}}
    \centerline{(b) Global Reverse}
\end{minipage}
\begin{minipage}{0.45\linewidth}
    \vspace{3pt}
    \centerline{\includegraphics[width=\textwidth]{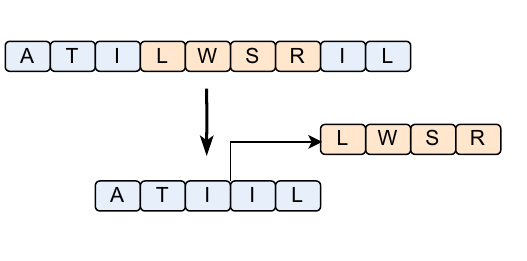}}
    \centerline{(c) Random Crop}
\end{minipage}
\begin{figure}[htb]
\caption{Illustrations of three protein augmentations, where unmodified and modified amino acids are marked in blue and orange.} \vspace{-0.5em}
\label{figure-appendix-extended}
\end{figure}

\section{Implementation Details}
Table.~\ref{table-training-details} shows the hyper-parameter of LSTM on five tasks, including EC, Sub, Bin, Fold and Yst. In terms of policy search hyper-parameters, we generally set the number of fine-tunes to 25 and allocate 5 epochs for each fine-tune stage. An exception is made for the subcellular localization task, where we adjust the number of fine-tunes to 20 and epochs per stage to 10. Consistently, across all tasks, we fix the number of sub-policies and operations at 4 and 2, respectively.

\begin{table}[hptb]
\begin{tabular}{llllll}
\toprule
Hyper-parameter & EC & Sub & \multicolumn{1}{c}{Bin} & \multicolumn{1}{c}{Fold} & \multicolumn{1}{c}{Yst} \\ \midrule
num of mlp layer & 3    & 2    & 2    & 2    & 2    \\
learning rate    & $5e^{-5}$ & $5e^{-5}$ & $5e^{-5}$ & $5e^{-5}$ & $5e^{-5}$ \\
batch size       & 8    & 32   & 32   & 16   & 32   \\
num epoch        & 95   & 50   & 75   & 25   & 50   \\
finetune num     & 25   & 20   & 25   & 25   & 25   \\
finetune epoch   & 5    & 10   & 5    & 5    & 5    \\
num subpolicy    & 4    & 4    & 4    & 4    & 4    \\
num operations   & 2    & 2    & 2    & 2    & 2    \\ \bottomrule
\end{tabular}
\caption{
Hyper-parameter settings on five protein-related tasks.
}
\label{table-training-details}
\end{table}

\end{document}